\documentclass{aa}

\usepackage{amsmath}
\usepackage{graphicx}
\usepackage{dcolumn}
\usepackage{bm}
\usepackage{epsfig}
\usepackage{amssymb,latexsym,mathrsfs}
\usepackage{enumitem}
\usepackage{color}
\usepackage{hyperref}
\usepackage{subcaption}
\usepackage{natbib}
\usepackage[utf8]{inputenc}
\usepackage[T1]{fontenc}

\makeatletter
\def\hlinewd#1{%
\noalign{\ifnum0=`}\fi\hrule \@height #1 \futurelet
\reserved@a\@xhline}
\makeatother

\usepackage[dvipsnames]{xcolor}

\newcommand{\msun}{M_{\odot}}
\newcommand{\be}{\begin{equation}}
\newcommand{\ee}{\end{equation}}
\newcommand{\bs}{\begin{split}}
\newcommand{\bea}{\begin{eqnarray}}
\newcommand{\eea}{\end{eqnarray}}
\definecolor{violet}{rgb}{0.62, 0.0, 1.0}

\begin{document}

    \title{Distinguishing Kilonovae from Binary Neutron Star and Neutron Star-Black Hole Mergers}

   \author{I. Gupta\inst{1,2,3}\corrauth{ishgupta@berkeley.edu}        
        \and Y. Bhoge\inst{4}\email{yugeshbhoge.edu@gmail.com}
        \and R. Kashyap\inst{4}\email{rahulkashyap@iitb.ac.in}
        \and M. Bhattacharya\inst{5,6,7,8}\email{}
        }

    \institute{
    Department of Physics, University of California, Berkeley, CA 94720, USA
    \and Department of Physics and Astronomy, Northwestern University, 2145 Sheridan Road, Evanston, IL 60208, USA
    \and Center for Interdisciplinary Exploration and Research in Astrophysics (CIERA), Northwestern University, 1800 Sherman Ave, Evanston, IL 60201, USA
    \and Department of Physics, Indian Institute of Technology Bombay, Mumbai 400076, India
    \and Department of Physics; Department of Astronomy \& Astrophysics; Center for Multimessenger Astrophysics
    \and Institute for Gravitation and the Cosmos, The Pennsylvania State University, University Park, PA 16802, USA
    \and Department of Physics, Wisconsin IceCube Particle Astrophysics Center, University of Wisconsin, Madison, WI 53703, USA
    \and Department of Astronomy, Astrophysics and Space Engineering, Indian Institute of Technology Indore, Simrol, MP 453552, India}

   \date{Received June XX, 20XX}

    \abstract
    {Kilonovae (KNe) are most informative when accompanied by a gravitational-wave signal, which can help identify the source as a binary neutron star (BNS) or a neutron star-black hole (NSBH) merger. However, future events will also be discovered serendipitously or through follow-up of other transients, without a confident identification of the progenitor.}
    {We ask whether the KN light curve alone can distinguish between these two progenitor channels.}
    {Using simulated BNS and NSBH populations together with semi-analytic light curve models, we compare their post-peak evolution across the optical $ugrizy$ bands, and quantify the separation between the two classes in each band with the area under the receiver-operating-characteristic curve (AUC).}
    {BNS and NSBH KNe populate distinct regions of the post-peak decline distribution, with BNS KNe fading faster in every band. The separation is cleanest in the blue $u$ and $g$ bands $5$ days after peak and in the redder $i$ band $10$ days after peak. Within $5$ days of peak, BNS KNe decline by $\gtrsim 6$ ($\gtrsim 4$) mag in $u$ ($g$) bands, whereas NSBH KNe fade by only $\sim 3$ ($\sim 1$) mag. Over $10$ days in $i$, NSBH KNe decline by $\sim 1$--$2$ mag against $\sim 3$--$6$ mag for BNS. We attribute this to the higher opacity of NSBH ejecta, which lengthens the photon-diffusion time and slows the decline in all bands, while a low opacity blue component drives the rapid early peak and decline of BNS KNe.}
    {Although the precise overlap is model-dependent, the qualitative separation persists across variations in the astrophysical population, the NS equation of state, and the controlled variation of ejecta model parameters, establishing the post-peak photometric decline as a viable EM-only diagnostic of whether a KN arose from a BNS or an NSBH merger.}


    \maketitle
    \nolinenumbers

\section{Introduction} \label{sec:intro}
Compact binaries containing at least one neutron star (NS) can be prolific multi-messenger sources. As these systems inspiral and merge, they radiate gravitational waves (GWs) detected by the LIGO-Virgo-KAGRA (LVK) detectors \citep{LIGOScientific:2025slb}. They can also eject neutron-rich material through tidal disruption or contact at merger. The resulting outflows power electromagnetic (EM) transients across the spectrum, including optical and near-infrared kilonovae (KNe) \citep{Li:1998bw,Kulkarni:2005jw,Metzger:2010sy}, gamma-ray bursts (GRBs) \citep{1989Natur.340..126E,1992ApJ...395L..83N}, and broadband afterglow emission \citep{Rees:1992zge,Wijers:1997xu,Sari:1997qe}. Whether and in what form these counterparts appear depend sensitively on the component masses, spin parameters, orbital configuration, and the NS equation of state (EOS), which together determine the amount, composition, geometry, and velocity of the ejecta.

The potential of joint GW-EM observations was established by the binary neutron star (BNS) event GW170817 \citep{LIGOScientific:2017vwq}, which produced a rich suite of EM counterparts from gamma-ray to radio \citep{LIGOScientific:2017ync,LIGOScientific:2017zic,DES:2017kbs,Cowperthwaite:2017dyu,Nicholl:2017ahq,Chornock:2017sdf,Margutti:2017cjl,Alexander:2017aly,Fong:2017ekk}. Combined with precise GW constraints, GW170817 catalyzed advances in transient modeling and multi-messenger inference \citep{Chornock:2017sdf,Villar:2017wcc,Radice:2017lry,Radice:2020ddv,Breschi:2021tbm,Gutierrez:2024pch}, established BNS mergers as the progenitors of at least some short-duration GRBs \citep{Rezzolla:2010fd,Rezzolla:2011da}, and confirmed compact object mergers as sites of heavy element ($r$-process) nucleosynthesis \citep{Lattimer:1974slx,1982ApL....22..143S, Thielemann:2017acv}.

In contrast, although neutron star–black hole (NSBH) mergers have been reported by LVK \citep{LIGOScientific:2021qlt}, and the sample has grown through the fourth observing run \citep{LIGOScientific:2025slb,LIGOScientific:2025pvj}, no EM counterpart has been confidently associated with any NSBH event despite extensive follow-up campaigns \citep[e.g.,][]{Anand:2020eyg,Ashkar:2020hxe,Kasliwal:2020wmy,Paterson:2020mmd,IceCube:2021ddq,Ronchini:2024lvb,Paek:2025dri}. This gap leaves open questions on whether typical NSBH mergers eject enough mass to power observable KNe~\citep{
Bhattacharya:2018lmw,Chattopadhyay:2022cnp}, which binary configurations are EM-bright \citep{Chattopadhyay:2020lff,Biscoveanu:2022iue,Gupta:2023evt}, whether NSBH mergers launch short or long-duration GRBs \citep{Zhu:2022kbt,Gottlieb:2023sja}, and how much these mergers contribute to $r$-process enrichment \citep{Saleem:2025bme}.

GW triggers provide critical information for EM counterpart search strategies. Signal-to-noise ratio accumulated during the inspiral phase can yield early-warning alerts with coarse localizations that enable rapid follow-up searches~\citep{Sachdev:2020lfd,Magee:2021xdx}. In addition, GW parameter estimation constrains component masses and spins of binaries \citep{LIGOScientific:2017vwq,LIGOScientific:2021qlt,LIGOScientific:2024elc,Gupta:2024bqn}, which informs the expected EM phenomenology as well as the plausible source class between BNS and NSBH mergers~\citep{Chaudhary:2023vec}. Planned upgrades, such as A+ \citep{Miller:2014kma,KAGRA:2013rdx}, and proposed next-generation facilities such as Cosmic Explorer \citep{LIGOScientific:2016wof,Reitze:2019iox,Evans:2023euw,Gupta:2023lga} and Einstein Telescope \citep{Punturo:2010zz,Hild:2010id,Branchesi:2023mws,ET:2025xjr} promise more frequent joint detections \citep{Zhu:2020ffa,Zhu:2021ram,Gupta:2022fwd}.
However, with near-term A+ sensitivities, the network reach\footnote{In \citet{Borhanian:2022czq} and \citet{Gupta:2023evt}, reach is defined as the redshift at which $50\%$ of mergers are detected.} for BNS (NSBH) events is $\sim\!0.1$ ($\sim\!0.2$) \citep{Borhanian:2022czq,Gupta:2023evt}, which is significantly lower than the redshifts at which state-of-the-art surveys such as the Vera C. Rubin Observatory \citep{LSST:2008ijt,Cowperthwaite:2018gmx,Andreoni:2021epw} and the Nancy Grace Roman Space Telescope \citep{Hounsell:2017ejq,Chase:2021ood,Andreoni:2023xlv} will detect KNe.

Efficient follow-up efforts also rely on GW-enabled sky localization and detector duty cycle. Two-detector baselines typically localize events to $\mathcal{O}(10^{2})$ deg$^{2}$ \citep{Berry:2014jja}, so simultaneous multi-detector GW observations are essential. Missed opportunities can result from single-detector triggers, as illustrated by the low mass NSBH candidate GW230529 \citep{LIGOScientific:2024elc}, which may have produced a KN \citep{Chandra:2024ila,Kunnumkai:2024qmw} but lacked a precise sky localization to enable targeted searches \citep{Ronchini:2024lvb,Paek:2025dri}. The second detected BNS merger, GW190425 \citep{LIGOScientific:2020aai}, also faced similar challenges \citep{Paek:2023xdc}.

Given that present and future EM facilities will continue to discover KNe serendipitously \citep{Andreoni:2023xlv}, and in association with other transients such as GRBs \citep{Tanvir:2013pia,Rastinejad:2022zbg}, we ask an intriguing question: \emph{in the absence of a GW detection, can a KN observation alone indicate whether the progenitor was a BNS or an NSBH merger?}

In this work, we model KN light curves for simulated BNS and NSBH populations with widely used semi-analytic prescriptions and numerical-relativity-calibrated fits. We assess whether these models predict observable differences between BNS and NSBH KNe that could enable EM-only source classification. We find that they do, as shown in Figure~\ref{fig:fid_diff_eos}. Motivated by the relationship between the intrinsic luminosity and decay rate of Type Ia supernovae~\citep{1993ApJ...413L.105P,SupernovaSearchTeam:1998fmf} and KNe~\citep{Kashyap2019-eo}, we introduce a distance-independent post-peak decline measure for KNe, defined as the difference between the peak KN magnitude and the magnitude after a fixed time in a given photometric band. We find that the post-peak decline in the blue ($u$ and $g$) bands provide a strong avenue for separating the two merger classes, with BNS KNe declining faster than NSBH KNe. This trend remains robust to variations in the assumptions of the astrophysical population, the NS EOS, and several parameters of the KN model. The same ordering is recovered in the red ($i$) band, where most NSBH KNe decline more slowly after peak than BNS KNe, although this distinction is more model-dependent. Because the $u$ band decline is steep and sensitive to the modeled photospheric temperature, we highlight the $g$ band as an equally discriminating, and, perhaps more robust, alternative. These results establish post-peak photometric decline as a promising EM-only probe of the progenitor, and motivate future efforts to revisit these regions of parameter space using comprehensive simulations.

The rest of the paper is organized as follows. Section~\ref{sec:kn_modeling} presents our semi-analytic framework for modeling BNS and NSBH KNe, along with its main assumptions and caveats. In Section~\ref{sec:comparison}, we compare the post-peak decline of the two merger classes and identify the bands and timescales where their KN emission differs most clearly. We then assess the robustness of these differences to variations in the astrophysical population priors (Section~\ref{subsec:comp_astropop}), NS EOS (Section~\ref{subsec:comp_eos}), and ejecta-model parameters (Section~\ref{subsec:comp_modpars}). We summarize our results and discuss their implications in Section~\ref{Sec:disc_conc}. Additional technical details are described in the Appendix.

\section{Kilonova Modeling} \label{sec:kn_modeling}

KNe are powered by the radioactive decay of $r$-process nuclei in neutron-rich ejecta, produced by the disruption of the NS in BNS and NSBH mergers. The decay of these radioactive heavy elements yields high-energy gamma rays that scatter off free electrons and convert to UV/optical/IR photons, generating transients in these bands. While both BNS and NSBH mergers can synthesize heavy elements, the electron fraction ($Y_e$) and lanthanide distribution can differ between ejecta components and between merger types (for example, see Table 1 in \citet{Ekanger:2023mde}). This directly impacts the effective opacity and the color evolution of the transient.

At early times, the ejecta is optically thick, so deposited radioactive energy is trapped, and the luminosity is regulated by photon diffusion. As the ejecta expands and rarefies, the optical depth decreases and radiation begins to escape, producing a rise to peak followed by a decline that increasingly tracks the declining radioactive power. The time to peak can be expressed by the geometric mean of two characteristic timescales: the expansion timescale, $\tau_{exp} = R/v$, and the diffusion timescale, $\tau_{diff} = \kappa M / \beta c R$ ~\citep{Metzger:2019zeh}. Here, $\kappa$ denotes the opacity of the ejecta component, $R$ is the length scale associated with the ejecta component, and $\beta = 13.8$ is a dimensionless constant that encodes the geometric structure of the ejecta~\citep{Chatzopoulos2012-km}. The ejecta mass $M$ and velocity $v$ are set by the binary system parameters and the nuclear EOS. In general, larger $M$ and $\kappa$ result in slower evolution of the light curve, while the latter also results in redder emission~\citep{Barnes:2013wka}.

We model KN emission within the Arnett–Chatzopoulos–Villar framework \citep{Arnett1982-wf,Chatzopoulos2012-km,Villar:2017wcc}, which maps ejecta masses, velocities, and opacities to bandwise light curves. For BNS mergers, we adopt the three-component prescription of \citet{Villar:2017wcc}, treating lanthanide-rich ``red'' component $(\kappa=10\,{\rm cm}^2\,{\rm g}^{-1})$, intermediate-opacity ``purple'' component $(\kappa=3\,{\rm cm}^2\,{\rm g}^{-1})$, and lanthanide-poor ``blue'' component $(\kappa=0.5\,{\rm cm}^2\,{\rm g}^{-1})$ as independent diffusion zones. For each BNS, the dynamical ejecta and remnant disk outflow masses and velocities are determined from the NS masses and EOS using numerical-relativity-calibrated fits from \citet{Nedora:2020qtd}. We then distribute the total ejecta among the three opacity components with fixed mass fractions $(f_{\rm red},f_{\rm purple},f_{\rm blue})=(0.2,0.6,0.2)$. The model is evolved separately for each component to obtain its luminosity as a function of time. The total bolometric light curve is the sum of the three components, and the corresponding band fluxes are obtained by summing up the component contributions in each filter.

In NSBH mergers, EM counterparts are produced only if the NS is tidally disrupted before reaching the BH’s innermost stable circular orbit (ISCO). Consequently, in addition to the NS mass and EOS, the existence and amount of ejecta depend strongly on the BH mass and spin. For sufficiently massive BHs, the NS typically plunges with little or no disruption. By contrast, a rapidly spinning BH with spin aligned to the orbital angular momentum reduces the effective ISCO radius and enhances tidal disruption, increasing both the remnant disk mass and the unbound outflows.
For disrupted NSBH systems, we adopt a two-component KN model consisting of a lanthanide-rich dynamical component and a remnant disk outflow (disk wind) component. We evolve each component as an independent one-zone diffusion region, characterized by its ejecta mass, expansion velocity, and effective opacity.
The dynamical ejecta is expected to be highly neutron-rich and, consequently, lanthanide-rich, while disk outflows typically synthesize lighter $r$-process nuclei, albeit with a lanthanide fraction that can vary with the disk properties and the central BH \citep{Just:2014fka,Wu:2016pnw,Lippuner:2017bfm}. Consistent with this picture, we adopt $Y_e \sim 0.05-0.1$ $(\kappa \sim 37\,{\rm cm}^2\,{\rm g}^{-1})$ for dynamical ejecta, and $Y_e \sim 0.1-0.5$ $(\kappa \sim 2-37\,{\rm cm}^2\,{\rm g}^{-1}, \mbox{see Appendix~\ref{appsubsubsec:nsbh_fits} for details})$ for disk outflows \citep{Ekanger:2023mde}.

For each NSBH binary, the ejecta-component masses and velocities are determined from numerical-relativity-calibrated fitting formulae that capture the dependence on mass ratio, BH spin, and NS compactness \citep{Foucart:2018rjc,Kruger:2020gig,Raaijmakers:2021slr}. We compute bolometric light curves for the dynamical and disk wind components within the same Arnett–Chatzopoulos–Villar framework used for BNS mergers, and obtain the total emission by summing the component contributions. The full set of fitting formulae for both BNS and NSBH mergers are provided in Appendix~\ref{appsec:det_kn_mod}.

To obtain bandwise light curves from the bolometric emission, we assume the ejecta radiates approximately as a blackbody. At each time step, we infer an effective photospheric radius and temperature from the luminosity and expansion, use these to construct the corresponding spectral energy distribution, and evaluate the spectral flux density at the effective wavelength of different photometric filters. We then convert the resulting fluxes to AB magnitudes, yielding bandwise light curves directly comparable to survey photometry. We generate light curves in the optical $ugrizy$ bands, the details of which are mentioned in Appendix~\ref{appsubsubsec:get_bandwise}.

There are several caveats to our KN modeling. First, we estimate ejecta masses and velocities using analytical fits to numerical relativity simulations. These fits can cause errors of at least $\mathcal{O}(10\%)$ in the predicted ejecta masses \citep{Kruger:2020gig,Foucart:2018rjc}, and in some regions of the parameter space, they can also be ill-conditioned \citep{Nedora:2020qtd}. Moreover, many of the underlying simulations follow the post-merger evolution for only tens of milliseconds and do not capture the full accretion evolution of the remnant disk and its long-lived outflows, largely due to computational cost. Some of these simulations also use approximate neutrino treatments rather than full neutrino transport, which can affect the predicted $Y_e$ distribution and, in turn, the opacities and heating rates that shape the KN light curves.

Second, the adopted prescription for radioactive heating introduces additional uncertainty \citep{Bulla:2022mwo,Sarin:2024tja}. In this work, we use a common functional form for the heating rate across both BNS and NSBH mergers. In reality, however, the isotopic abundance pattern can differ between the ejecta of the two merger classes, leading to differences in both the normalization and the temporal evolution of the heating. Third, we assume isotropic emission and neglect viewing-angle effects arising from aspherical ejecta geometry and composition gradients. In realistic systems, both the angular distribution of the ejecta and angle-dependent opacities can imprint substantial anisotropy on the luminosity and color evolution of the transient \citep{Klion2020-cx,Heinzel2020-co,Darbha2020-nz,Shingles2023-az}.

While we test the robustness of our conclusions against several modeling assumptions (see, for e.g., Section~\ref{subsec:comp_modpars} and Appendix~\ref{appsubsubsec:bns_fits}), a more definitive assessment will ultimately require a dedicated (albeit computationally prohibitive) set of high-resolution simulations with self-consistent treatment of nucleosynthesis, together with full radiative transfer. This work shows that, within widely used semi-analytic prescriptions, BNS and NSBH KNe are distinguishable, and motivates future efforts to study these unexplored regions of parameter space with more sophisticated models.

\section{Differentiating between BNS and NSBH Kilonovae}
\label{sec:comparison}

To quantify how KNe from BNS and NSBH mergers differ, we simulate populations of both classes and generate bandwise light curves using the framework summarized in Section~\ref{sec:kn_modeling} and delineated in Appendix~\ref{appsec:det_kn_mod}. For the fiducial BNS population, we draw the primary mass $m_1$ uniformly from $\mathcal{U}(1,2.5)\,\msun$ and the secondary mass $m_2\in\mathcal{U}(1,m_1)\,\msun$. For NSBH binaries, we sample the BH mass $m_{\rm BH}\in\mathcal{U}(2.5,12)\,\msun$, the NS mass $m_{\rm NS}\in\mathcal{U}(1,2.5)\,\msun$, and the dimensionless BH spin $a_{\rm BH}\in\mathcal{U}(-0.75,0.75)$. All systems are placed at a fixed luminosity distance of 200 Mpc. To capture the dependence on EOS, we repeat the analysis for three representative choices: SLy \citep{Douchin2001-to}, APR4 \citep{Akmal:1998cf} and DD2 \citep{Banik2014-mb}, spanning soft to stiff behavior. Although our population permits NS masses up to $2.5\,\msun$ (near the maximum supported by DD2), light curves are generated only for NS masses consistent with the EOS under consideration. We also enforce $m_{\rm BH,min}=2.5\,\msun$, deliberately targeting the regime in which a low mass BH and a high mass NS can be difficult to distinguish using GW tidal information alone~\citep{Golomb:2024mmt,Dhani2025-rq}. In such cases, systematic differences in the KN photometric evolution offer an EM-only route to source classification.

Our strategy is to statistically compare KNe from BNS and NSBH mergers using distance-independent observables that are directly measurable from survey photometry. We characterize each light curve by its peak AB magnitude in a given band, $m_{\rm AB}^{\rm peak}$, and its post-peak decline over a fixed time,
$m_{\rm AB}^{\rm peak}-m_{\rm AB}^{\rm peak+\Delta t}$, with $\Delta t = 1,2,5,10,$ and 15 days. The latter parameter captures the decline rate while remaining insensitive to luminosity distance. This approach is inspired by \citet{Kashyap2019-eo}, who showed that, similar to Type IA supernovae, the bolometric decline rate for BNS KNe correlates with their intrinsic bolometric luminosity, suggesting that measuring post-peak decline can help in standardizing KNe.

The temporal evolution of these distributions across all modeled bands and time windows is shown in Figure~\ref{appfig:all_bands} (Appendix~\ref{appsec:all_bands}). To quantify how cleanly the two merger classes separate in a given band and at a given $\Delta t$, we summarize each pair of distributions by the area under the receiver-operating-characteristic curve (AUC)~\citep{HanleyMcNeil1982}. For our case, the AUC is equal to the probability that a randomly chosen NSBH KN has a slower post-peak decline, i.e., fades more slowly, than a randomly chosen BNS KN (see Appendix~\ref{appsec:all_bands}). A value of $0.5$ indicates indistinguishable populations, while $1$ indicates perfect separation between the two distributions.

\begin{figure}
    \centering
    \includegraphics[width=0.49\textwidth]{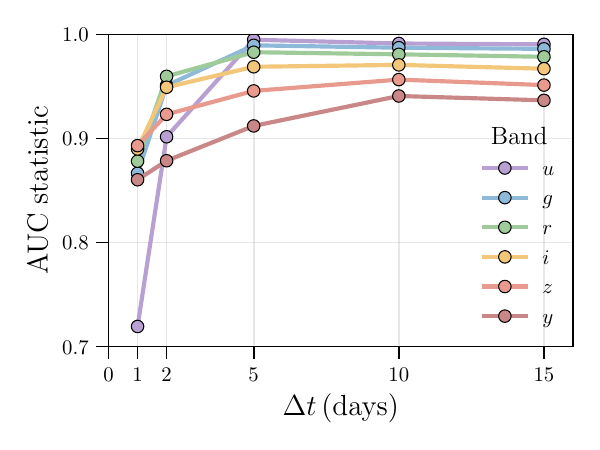}
    \caption{Area under the receiver-operating-characteristic curve (AUC) for the BNS--NSBH post-peak decline distributions, as a function of the time after peak $\Delta t$, for each of the $ugrizy$ bands. The reported AUC statistic equals the probability that a randomly chosen NSBH KN fades more slowly than a randomly chosen BNS KN, hence representing the distinguishability between the two distributions. ${\rm AUC} = 0.5$ indicates indistinguishable populations, while ${\rm AUC}=1$ indicates perfect separation. The bluer bands are the weakest discriminators at early times but overtake the redder bands by $\Delta t \simeq 5$ d, where the $u$ and $g$ bands reach $\gtrsim 0.99$. The reddest bands ($z$, $y$) continue to improve until $\Delta t = 10$ d.}
    \label{fig:auc_time}
\end{figure}

\begin{figure*}
    \centering
    \includegraphics[trim=0.3cm 0.3cm 0.3cm 0.3cm,
  clip,width=0.85\textwidth]{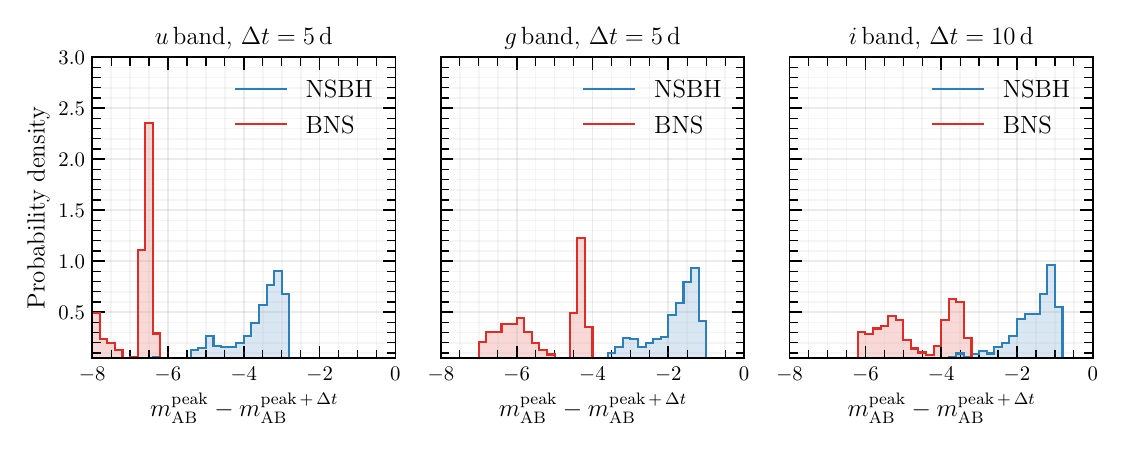}
  \caption{Distribution of the post-peak decline $m_{\rm AB}^{\rm peak} - m_{\rm AB}^{{\rm peak}+\Delta t}$ for KNe from BNS (red) and NSBH (blue) mergers. Left: $u$ band with $\Delta t = 5$ d. Middle: $g$ band with $\Delta t = 5$ d. Right: $i$ band with $\Delta t = 10$ d. Each histogram combines samples from all three NS EOS considered. The two populations are well separated in all three bands, with BNS KNe declining faster than NSBH KNe. The contrast is cleanest in the $u$ and $g$ bands, with slight overlap in the BNS and NSBH distributions in the $i$ band.
} \label{fig:fid_diff_eos}
\end{figure*}

Figure~\ref{fig:auc_time} shows the AUC as a function of $\Delta t$ for each band, considering the fiducial BNS and NSBH populations and combining samples across the three EOSs. The separation between the BNS and NSBH post-peak decline distributions improves with time and then saturates. The bluer bands ($u$, $g$) reach their maximum AUC $(\gtrsim 0.99)$, and hence, distinguishability, by $\Delta t \simeq 5$ d, while the slower-evolving reddest bands ($z$, $y$) continue to improve until $\Delta t \simeq 10$ d. Because the AUC rises and saturates at band-dependent times, reaching a maximum in the blue
bands by $\sim 5$ d and in the reddest bands only by $\sim 10$ d, its time dependence can inform follow-up strategy, indicating which band and which
post-peak epoch most cleanly separate the two progenitor channels for a given campaign.

Guided by these trends, we select two representative diagnostics for identifying the source of the KN: the $u$ band decline $5$ days after peak and the $i$ band decline $10$ days after peak. The $u$ and $g$ bands provide the largest and cleanest separation (AUC $\gtrsim 0.99$) between the two merger classes. On the other hand, $i$ band represents the redder optical bands, and because the redder ejecta is slower-evolving, we probe the decline in the $i$ band magnitude $10$ d after the peak. Figure~\ref{fig:fid_diff_eos} shows the post-peak decline distributions in the
$u$, $g$, and $i$ bands for the fiducial BNS and NSBH populations, combining samples across the three EOSs. In all three filters, the two merger classes populate distinct regions of the post-peak decline parameter space, with BNS KNe declining faster than NSBH KNe.

The contrast is largest in the blue bands: BNS events dim by $\gtrsim 6$ $(\gtrsim 4)$ mag in the $u$ $(g)$ band within $5$ days of peak, whereas NSBH events dim by only $\sim 3$ $(\sim 1)$ mag over the same interval. In the $i$ band, the larger masses in high opacity components increase the characteristic diffusion time and sustain red emission for longer, producing a slower decline ($\sim 1-2$ mag over $10$ d) for most NSBH events with non-negligible ejecta than for BNS KNe ($\sim 3-6$ mag).

These behaviors can be understood in terms of the ejecta properties that control the diffusion time and color evolution: the ejecta mass, velocity, and effective opacity. The decline rate is set by the photon diffusion timescale, $\tau_{\rm diff}$. The two merger classes differ systematically in both the mass and the opacity of their ejecta. For DD2, the BNS ejecta masses span $\sim 0.003$--$0.01\,\msun$ in the red and blue components and $\sim 0.01$--$0.03\,\msun$ in the purple component. Thus, the low opacity blue component ($\kappa\simeq 0.5~{\rm cm^2\,g^{-1}}$) has a short diffusion time: it dominates the early blue emission, peaks quickly, and then fades rapidly, while the higher-opacity purple and red components dominate progressively later times. In the NSBH model, only a subset of binaries undergo NS disruption outside ISCO and produce non-negligible ejecta. For those systems, the ejecta is typically more neutron-rich and opaque. For DD2, the disrupted NSBH systems in our population have median dynamical ejecta and disk wind masses of $\sim 0.025\,\msun$ and $\sim 0.013\,\msun$, respectively, with both components assigned substantially
larger effective opacities than the BNS blue component. The larger masses and higher opacities lengthen the diffusion time and broaden the light curve, so
NSBH KNe fade more slowly than BNS KNe in the considered bands.

Note that the very steep decline for BNSs in the $u$ band that makes it such a clean discriminator also pushes the band far onto the Wien tail of the cooling photosphere, where the predicted flux is most sensitive to our blackbody and recombination temperature-floor treatment (Appendix~\ref{appsubsubsec:get_bandwise}). Moreover, the source may be too faint in the $u$ band $5$ days after peak to be observable. The $g$ band provides a pragmatic alternative. It exhibits similar source distinguishability (AUC $\simeq 0.99$ at $5$ d) with a milder, more modeling-robust decline (for instance, see Figure~\ref{appfig:comp-gw170817} in Appendix~\ref{appsubsubsec:comp_gw170817}), and can be used interchangeably with the $u$ band diagnostic when it is the more practical choice for a given instrument.

Beyond modeling systematics, the degree of separation between BNS and NSBH KNe can also depend on the assumptions used to construct the underlying binary populations and determine ejecta properties. In the remainder of this section, we test the robustness of the inferred KN distinguishability under variations of these inputs. In Section~\ref{subsec:comp_astropop}, we repeat the analysis for alternative, astrophysically motivated BNS and NSBH populations and check how the distinguishability of the post-peak decline distribution changes. In Section~\ref{subsec:comp_eos}, we isolate EOS effects by comparing the separation in the post-peak decline distributions across the three NS EOS choices. Finally, in Section~\ref{subsec:comp_modpars}, we vary the assumed $Y_e$ (and, hence, the opacity) for the NSBH dynamical ejecta and disk wind components, and mass and velocity of the three BNS ejecta components, and reassess the resulting light curve differences.

\subsection{Impact of astrophysical population} \label{subsec:comp_astropop}

\begin{figure*}
    \centering
    \includegraphics[trim=0.3cm 0.3cm 0.3cm 0.3cm,
  clip,width=0.85\textwidth]{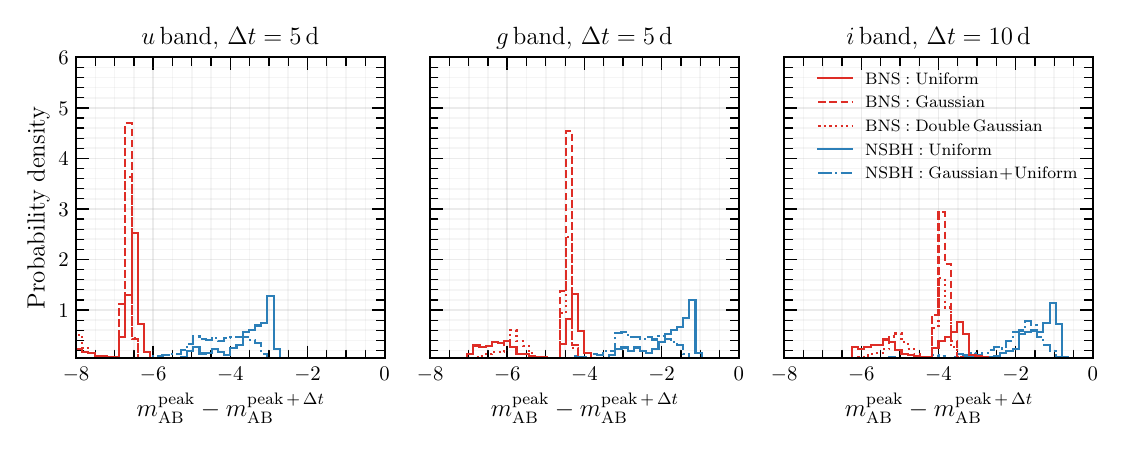}
    \caption{Same as Figure~\ref{fig:fid_diff_eos}, but while varying the astrophysical population priors for BNS and NSBH mergers and fixing the EOS to DD2. The separation in the $u$ band is largely insensitive to the adopted population. In the $g$ and $i$ bands, the ``Gaussian + Uniform'' NSBH prior shows slightly higher overlap with the BNS distributions.}
  \label{fig:var_astro}
\end{figure*}

To assess whether our result depends on population assumptions, we repeat the analysis using additional, astrophysically motivated priors for BNS and NSBH masses and spins. The adopted mass distributions are summarized in Table~\ref{tab:pop_pars}. For BNS systems, we consider a single Gaussian motivated by Galactic double NS measurements \citep{Farrow:2019xnc}, and a double-Gaussian mixture that incorporates additional pulsar constraints \citep{Antoniadis:2016hxz,Ozel:2016oaf} and follows \citet{Gupta:2023lga}. For NSBH systems, we adopt a ``Gaussian + Uniform'' prescription in which the BH mass is drawn from a Gaussian distribution motivated by the detected NSBH sample \citep{Biscoveanu:2022iue,LIGOScientific:2024elc}, while the NS mass remains uniform. We assume NSs to be non-spinning, and draw BH spins from $a_{\rm BH}\sim \mathcal{N}(0,0.2)$, to reflect the preference for low spins suggested by current detections~\citep{Biscoveanu:2022iue}. The NS EOS is fixed to DD2.

\begin{table}[h]
    \centering
    \begin{tabular}{l|cc}
    \hline
    Population & $m_{1}\ (\msun)$ & $m_{2}\ (\msun)$\\
    \hline
    \multicolumn{3}{c}{BNS}\\
    \hline
    Uniform & $\mathcal{U}(1,2.5)$ & $\mathcal{U}(1,m_1)$\\
    Gaussian & \multicolumn{2}{c}{$\mathcal{N}(1.33,0.9)$}\\
    Double Gaussian & \multicolumn{2}{c}{$0.64\!\times\!\mathcal{N}(1.33,0.9) + 0.36\!\times\!\mathcal{N}(1.8,0.3)$}\\
    \hline
    \multicolumn{3}{c}{NSBH}\\
    \hline
    Uniform & $\mathcal{U}(2.5,12)$ & $\mathcal{U}(1,2.5)$\\
    Gaussian\!+\!Uniform & $\mathcal{N}(5,1)$ & $\mathcal{U}(1,2.5)$ \\
    \hline
    \end{tabular}
    \caption{Parameters describing the BNS and NSBH mass distributions considered in Section~\ref{subsec:comp_astropop}. $\mathcal{U}(a,b)$ denotes a uniform distribution between $a$ and $b$, and $\mathcal{N}(\mu,\sigma)$ denotes a Gaussian distribution with mean $\mu$ and standard deviation $\sigma$. For all distributions, we enforce $m_1 > m_2$.}
    \label{tab:pop_pars}
\end{table}

Figure~\ref{fig:var_astro} compares the resulting post-peak decline distributions with those obtained for the fiducial ``Uniform'' populations. As expected, the ``Gaussian'' populations generally produce narrower distributions than their ``Uniform'' counterparts, since they occupy a more restricted region of the parameter space. Overall, however, the separation between BNS and NSBH KNe persists under these changes in population priors. This robustness is especially evident in the $u$ band, where the BNS population continues to occupy the rapidly declining region with little overlap with the more slowly declining NSBH events.

For BNS mergers, the ``Gaussian'' and ``Double Gaussian'' populations contain fewer systems with $m_{\rm NS}\lesssim 1.2\,\msun$ and $m_{\rm NS}\gtrsim 2\,\msun$. As a result, their $i$ band distributions lack both the fastest-declining $(\gtrsim\!6$ mag) and the slowest-declining $(\lesssim\!3.5$ mag) tails seen in the fiducial ``Uniform'' population (cf. Figure~\ref{appfig:bin_corner} in Appendix~\ref{appsec:bin_prop_ejec}). For NSBH mergers, the ``Gaussian + Uniform'' population shows somewhat greater overlap with the ``Uniform'' BNS distribution in the $g$ and $i$ bands, reflecting a shift toward faster post-peak decay. This behavior arises because the ``Gaussian + Uniform'' prescription concentrates BH masses around $\sim 5\,\msun$. Relative to the fiducial ``Uniform'' prior, this increases the fraction of moderately mass-symmetric systems, but decreases the abundance of the most ejecta-producing binaries: near-equal-mass NSBH systems with $m_{\rm BH}/m_{\rm NS}\lesssim 2$ and large aligned BH spins. The ``Uniform'' NSBH prior, although it includes many highly mass-asymmetric binaries that do not disrupt, still retains more of this strongly ejecta-producing tail. Consequently, among systems that produce non-negligible ejecta, the ``Gaussian + Uniform'' population is weighted toward smaller ejecta masses and shorter diffusion timescales, leading, on average, to faster fading in the $u$, $g$, and $i$ bands.

\subsection{Impact of the NS equation of state}
\label{subsec:comp_eos}

\begin{figure*}
    \centering
    \includegraphics[trim=0.3cm 0.3cm 0.3cm 0.3cm,
  clip,width=0.85\textwidth]{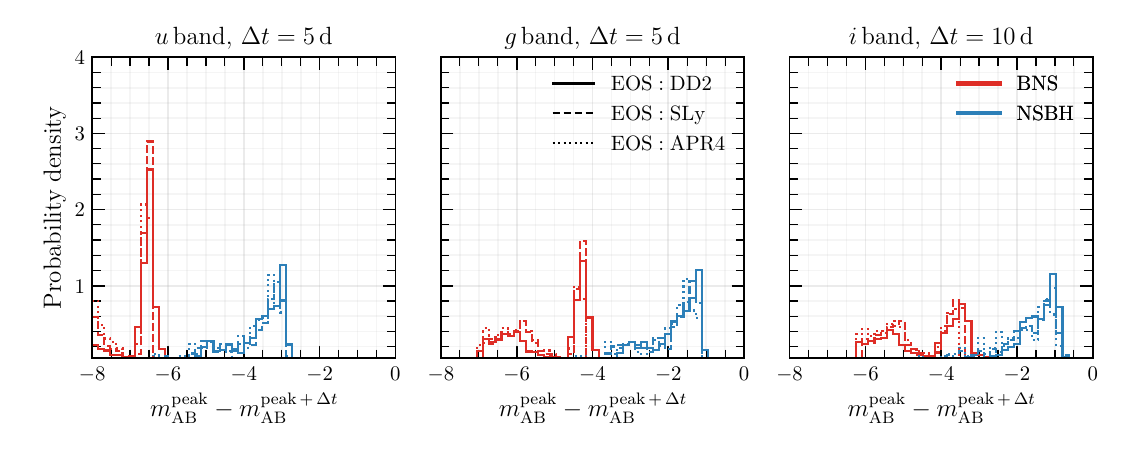}
    \caption{Same as Figure~\ref{fig:fid_diff_eos}, but shown separately for each EOS. Varying the EOS produces only a small shift in the post-peak decline distributions and does not significantly change the separability between BNS and NSBH KNe in any of the bands.}
    \label{fig:var_eos}
\end{figure*}

The EOS sets the NS compactness, which is an important parameter for tidal disruption and mass ejection. More compact NSs are harder to disrupt and, when disrupted, tend to produce less ejecta. Since the KN decline rate is sensitive to ejecta mass, EOS variations can alter the post-peak decline distribution. Hence, we repeat the analysis for the fiducial ``Uniform'' BNS and NSBH populations under each of the three EOSs and reassess the BNS--NSBH separability.

Among our EOS choices, APR4 generally yields the most compact (softest) NSs, followed by SLy, while DD2 produces the least compact (stiffest) NSs. The resulting trends in ejecta production follow this ordering. For BNS mergers, DD2 yields the largest ejecta masses on average, followed by SLy and then APR4. For NSBH mergers, the EOS also affects the disruption fraction, because compactness enters directly into the criterion for disruption outside the BH ISCO in the calibrated fits. In our simulated NSBH population, the fraction of systems with non-negligible ejecta is $\sim 11\%$ for DD2, $\sim 6\%$ for SLy, and $\sim 4\%$ for APR4.

Figure~\ref{fig:var_eos} shows the resulting post-peak decline distributions in the $u$, $g$, and $i$ bands for each EOS. As we move from DD2 to APR4, both the BNS and NSBH populations shift mildly toward faster KN decay in both bands, consistent with smaller ejecta masses and shorter diffusion times for more compact stars. Importantly, this shift is subdominant to the separation between merger classes. While the overlap between populations is slightly higher for APR4, the $u$ and $g$ bands continue to show significant distinguishability between BNS and NSBH KNe, while the $i$ band separation persists with comparable overlap across EOS choices. We conclude that, within our modeling assumptions, the proposed post-peak decline criteria are reasonably robust to EOS uncertainty.

\subsection{Impact of model parameters}
\label{subsec:comp_modpars}

Finally, we examine how uncertainties in the ejecta composition, opacity, and characteristic velocity affect the BNS--NSBH separation in the post-peak decline distribution. Throughout this analysis, we keep the astrophysical population fixed to the fiducial ``Uniform'' prior and adopt DD2 as the EOS for both BNS and NSBH systems.

For BNS mergers, we vary the allocation of the total unbound ejecta mass among the blue, purple, and red components from the fiducial fractions $f_m=\{0.2,0.6,0.2\}$ to $f_m=\{0.26,0.60,0.14\}$. The latter choice is motivated by the component-wise ejecta masses inferred for the KN associated with GW170817 \citep{Villar:2017wcc}. In the fiducial model, all three components are assigned the same characteristic velocity. In the alternate model, guided by the inferred component-wise ejecta velocities for GW170817~\citep{Villar:2017wcc}, we scale the velocities of the blue, purple, and red components as $\{1,0.6,0.5\}$ times the ejecta velocity inferred from the numerical-relativity-calibrated fits (see Appendix~\ref{appsubsubsec:bns_fits}).

For NSBH mergers, we instead vary the electron fraction, and hence the opacity, of the ejecta components. As discussed in Section~\ref{sec:kn_modeling}, the dynamical ejecta is expected to be highly neutron-rich and have low $Y_e$, while the disk outflow is expected to be less neutron-rich and have lower opacity. We vary the electron fraction of the dynamical ejecta from $Y_e=0.1$ to $Y_e=0.05$, and that of the disk wind from the fiducial value $Y_e=0.3$ to $Y_e=\{0.1,0.5\}$ \citep{Ekanger:2023mde}. Using the fits to the opacity data from \citet{Tanaka:2019iqp}, these values correspond to $\kappa=\{36.95,\,36.82,\,3.59,\,1.98\}\ {\rm cm}^2\,{\rm g}^{-1}$ for $Y_e=\{0.05,\,0.1,\,0.3,\,0.5\}$, respectively (see Appendix~\ref{appsubsubsec:nsbh_fits}). Since the fitted opacity changes only minimally between $Y_e=0.05$ and $0.1$, varying the composition of the dynamical ejecta within this range has a negligible impact on the resulting KN light curves. Thus, we focus on the effect of varying the disk wind composition.

Figure~\ref{fig:var_Ye} shows the resulting changes in the post-peak decline distributions. For BNS mergers, the reduced velocities of the purple and red components in the alternate model slow the expansion, leading to less rarefied ejecta and more slowly evolving emission in the $i$ band. By contrast, the $u$ and $g$ band distributions are only weakly affected, since they are dominated by the blue component, whose velocity remains unchanged and whose emission peaks and decays on short timescales.

For NSBH mergers, the disk wind electron fraction strongly alters the color evolution through the opacity. The $Y_e^{\rm disk}=0.5$ case corresponds to the lowest opacity disk wind and, therefore, to a bluer KN. In this case, the lower opacity shortens the diffusion time, causing both the light curves in all three bands to decay more rapidly than in the fiducial $Y_e^{\rm disk}=0.3$ model. Because BNS KNe are themselves the faster-declining population, this shift moves the NSBH distribution toward the BNS one, increasing their overlap in the $i$ band and reducing the separation in the $u$ and $g$ bands. However, the two merger classes are still largely distinguishable in all bands.
The opposite trend is seen for $Y_e^{\rm disk}=0.1$, which corresponds to a much higher opacity disk wind and, consequently, a redder KN. In all three bands, this produces the slowest declining light curves and the largest separation from BNS distributions.

Overall, these variations show that the precise location of the BNS and NSBH populations in the post-peak decline distribution depends on the ejecta modeling, particularly through the opacities and velocities assigned to the components. Nevertheless, the qualitative picture remains the same. Across the model variations considered here, BNS and NSBH KNe continue to populate systematically different post-peak decline distributions, with the separation in the bluer bands remaining especially robust, while the redder $i$ band distinguishability is more sensitive to assumptions about the composition and velocity structure of the ejecta. We conclude that, although the precise degree of separation in the post-peak decline distribution is model-dependent, the existence of observationally accessible differences between BNS and NSBH KNe is likely not an artifact of a single fiducial prescription.

\section{Conclusions}
\label{Sec:disc_conc}

\begin{figure*}
    \centering
    \includegraphics[trim=0.3cm 0.3cm 0.3cm 0.3cm,
  clip,width=0.85\textwidth]{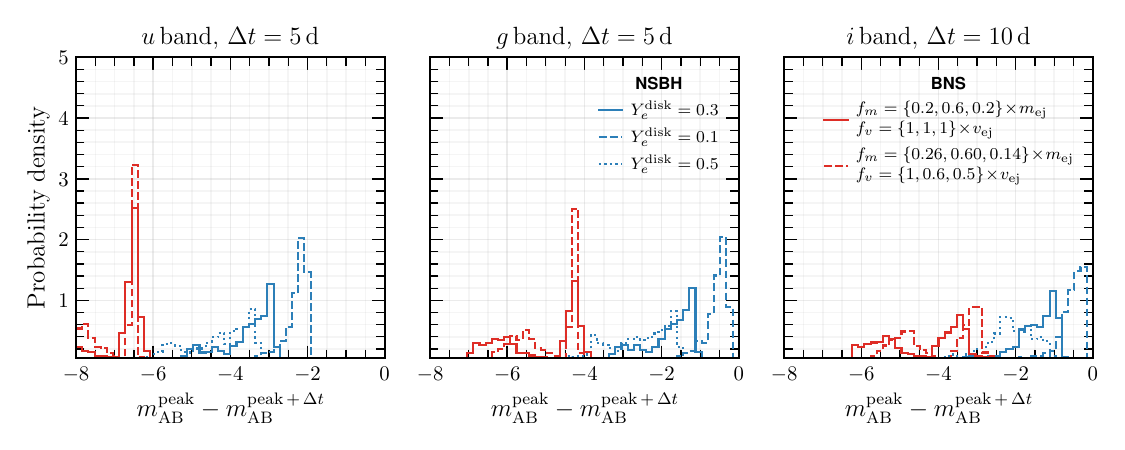}
  \caption{Same as Figure~\ref{fig:fid_diff_eos}, but illustrating the impact of variations in the ejecta-model assumptions for the fiducial ``Uniform'' populations with the DD2 EOS. For BNS mergers, we vary the allocation of the total unbound ejecta mass, $m_{\rm ej}$, among the blue, purple, and red components from the fiducial fractions $f_m=\{0.2,0.6,0.2\}$ to $f_m=\{0.26,0.60,0.14\}$, and also modify the characteristic ejecta velocities, $v_{\rm ej}$, of these components. For NSBH mergers, we instead vary the electron fraction of the disk wind ejecta, changing the fiducial value $Y_e=0.3$ (corresponding to $\kappa\sim 3.6\,{\rm cm}^2\,{\rm g}^{-1}$) to $Y_e=0.1$ ($\kappa\sim 37\,{\rm cm}^2\,{\rm g}^{-1}$) and $Y_e=0.5$ ($\kappa\sim 2\,{\rm cm}^2\,{\rm g}^{-1}$).}
  \label{fig:var_Ye}
\end{figure*}

Kilonovae (KNe) are optical and near-infrared transients powered by the radioactive decay of freshly synthesized $r$-process nuclei in neutron-rich ejecta. They were first identified as counterparts to some short GRBs, pointing to an origin in compact binary mergers. In particular, the merger of two NSs, or the tidal disruption of a NS by a BH, can eject neutron-rich matter under conditions favorable for heavy-element nucleosynthesis. This picture was confirmed by GW170817, a BNS merger that was accompanied by both a short GRB and a KN. However, no comparably secure association has yet been established for an NSBH merger.

A practical consequence of not detecting the associated GW signal is that the progenitor class of a KN cannot be inferred. This is especially relevant as optical surveys, like the Vera C. Rubin Observatory and the Nancy Grace Roman Space Telescope, begin to probe beyond the effective multi-messenger reach of the present GW networks, and because even GW-detected systems can remain difficult to classify in parts of the BNS--NSBH parameter space. This motivates the central question of this work: can one infer the progenitor class of a KN from the detected EM signal alone? Using simulated populations and semi-analytic light curve models calibrated to numerical simulations, we find that BNS and NSBH mergers do exhibit statistically different photometric evolution, opening a path toward EM-only source classification.

To quantify these differences, we compare BNS and NSBH KNe across the optical $ugrizy$ bands using the post-peak decline distributions, constructed from the peak AB magnitude in a given band and the change in magnitude over a fixed interval after peak. This quantity is directly measurable from survey photometry, and the post-peak decay itself is independent of luminosity distance. We summarize the separation between the two classes in each band with
the AUC statistic ($0.5$ represents no model distinguishability whereas $1$ represents perfect separation), which reaches $\sim 0.99$ in the blue ($u$, $g$) and exceeds $\sim 0.95$ in $r$ and $i$ by $5$ days after peak (c.f. Figure~\ref{fig:auc_time}). Thus, the separation is cleanest in the blue $u$ and $g$ bands $5$ days after peak: in the $u$ ($g$) band, typical BNS KNe fade by $\gtrsim 6$ ($\gtrsim 4$) mag within $5$ days of peak, whereas NSBH KNe fade by only $\sim 3$ ($\sim 1$) mag over the same interval. The same ordering holds in the $i$ band, where NSBH KNe generally decline more slowly ($\sim 1-2$ mag) over $10$ days than BNS KNe ($\sim 3-6$ mag) (c.f. Figure~\ref{fig:fid_diff_eos}). We note that capturing the peak in any EM band is contingent on early localization and timely follow-up. In this light, Figure~\ref{fig:auc_time} can be used to identify the optimal band, i.e., the one that maximizes the AUC while allowing its peak to be located with minimal uncertainty.

This behavior follows from the ejecta properties that control the KN timescale and color evolution. In BNS mergers, the early blue emission receives a
substantial contribution from a low opacity component with a short diffusion time, which peaks and fades quickly and drives a steep early decline in the $u$ band. Ejecta-producing NSBH mergers, by contrast, typically contain more neutron-rich material with larger effective opacities in both the dynamical and disk wind ejecta. The resulting longer diffusion times broaden the light curve and slow the post-peak decline in every band, including the blue. The same larger, higher-opacity ejecta shift more of the emission to later times and redder wavelengths, sustaining a slow $i$ band decline. Because the $u$ band decline is steep and pushes the band onto the Wien tail of the cooling photosphere, where the predicted flux is low and most sensitive to the
photospheric-temperature treatment, we identify the $g$ band as an equally discriminating, and perhaps more robust, alternative.

We test how robust this separation in the post-peak decline distribution is to several assumptions. First, changing the astrophysical population priors does not erase the difference between BNS and NSBH KNe. Our fiducial ``Uniform'' populations are already intentionally broad and agnostic, spanning a wide range of masses and spins, yet they still show clear separation, particularly in the bluer $u$ and $g$ bands (c.f. Figure~\ref{fig:var_astro}). Adopting more astrophysically motivated priors generally narrows the distributions and, over much of the parameter space, preserves or even strengthens the contrast between BNS and NSBH systems.

Second, we vary the NS EOS, which changes the NS compactness and, hence, the amount of ejecta produced. This effect is especially important for NSBH mergers, where more compact NSs are less likely to disrupt outside the BH's innermost stable circular orbit and less likely to produce an observable KN. Considering NSBH systems that produce non-negligible ejecta, the relative placement of the BNS and NSBH populations in the post-peak decline distribution changes only slightly (c.f. Figure~\ref{fig:var_eos}). Softer equations of state, corresponding to more compact NSs, generally yield smaller ejecta masses and faster evolution in all three bands, but they do not significantly alter the distinguishability between the two merger classes.

Third, we varied model parameters that directly control the KN color and diffusion timescale, including the mass and velocity of the BNS ejecta components, and the opacity of the NSBH disk wind. These changes shift the distributions in ways that are broadly consistent with physical expectations. Lower opacity NSBH disk winds shorten the diffusion time, producing more rapidly declining light curves in both the blue and red bands, while higher opacity winds lengthen the diffusion time and sustain slowly declining, redder light curves for longer (c.f. Figure~\ref{fig:var_Ye}). Likewise, assigning lower velocities to the BNS's purple and red ejecta components causes ejecta to expand and rarefy more slowly, increasing the effective diffusion timescale and slowing the post-peak decay in the $i$ band. Even so, these variations do not qualitatively alter the main result: across the model space explored here, BNS and NSBH KNe continue to occupy systematically different regions in the post-peak decline distribution.

These conclusions should be interpreted with appropriate caution. Our modeling relies on semi-analytic light curve prescriptions and simulation-calibrated ejecta fits, inheriting uncertainties from the underlying numerical simulations, fitting routines, heating rate prescriptions, opacity estimates, and neglecting viewing-angle dependence and detailed radiative transfer. In particular, the steep $u$ band decline places that band on the Wien tail of the modeled photosphere, making it the most sensitive to the adopted blackbody temperature and recombination-floor prescription. Hence, we also emphasize the more robust $g$ band observations as an alternative to the $u$ band diagnostic. We explicitly tested the stability of our results against several assumptions, but a definitive EM-only classification framework will ultimately require a large suite of high-resolution simulations with self-consistent neutrino transport, composition-dependent heating, and full radiative transfer calculations. Even with these caveats, the central result is clear: within widely-used semi-analytic models, BNS and NSBH KNe exhibit systematically different photometric evolution. The persistence of this separation across population priors, NS EOS choices, and ejecta-model variations shows that the prospect of EM-only source classification is not tied to any one particular prescription, but instead reflects a broader physical trend that merits detailed follow-up with state-of-the-art simulations.

If these trends persist under more sophisticated modeling, their implications are significant. Current KN datasets are generally too sparse to apply this framework, especially for events whose light curves have not informed the semi-analytic prescriptions used here. That situation may change rapidly with facilities such as Rubin and Roman, which are expected to discover larger samples of KNe, including better-sampled counterparts to short GRBs and GW events. In that regime, the post-peak decline behavior identified here could help infer the likely nature of the underlying binary even when no GW information is available. The same is relevant when a GW signal is present but not decisive. At current detector sensitivities, low mass BHs can be difficult to distinguish from heavy NSs, and such systems are explicitly included in our broad NSBH population. Our results indicate that KNe of such low mass NSBHs can still differ from those of BNS mergers (c.f. Appendix~\ref{appsec:bin_prop_ejec}). If confirmed, this would provide an additional route to source classification, with direct implications for the maximum NS mass and the NS EOS. More broadly, robust identification of BNS and NSBH KNe would also help address open questions about the relative contribution of different compact binary channels to Galactic $r$-process enrichment, and about whether NSBH mergers preferentially power short- or long-duration GRBs.

The natural next step is to test these predictions using more complex ejecta and heating prescriptions, explicit radiative transfer, and viewing-angle dependence. Given the qualitatively different evolution we find in the blue and red optical bands, it will also be valuable to extend this analysis to ultraviolet and infrared wavelengths, where the contrast between BNS and NSBH KNe may be even more pronounced. Ultimately, the strongest assessment will require an apples-to-apples comparison of BNS and NSBH mergers using high-resolution general relativistic magnetohydrodynamic simulations and consistent post-processing pipelines to determine whether the separation we identify in the post-peak decline distribution persists in more complete models. If it does, KN light curves will offer a new way to connect EM transients to their progenitors and, in turn, to the astrophysics of compact binary mergers.

\section*{Contributions}
IG led the project and carried out the large-scale simulations for BNS and NSBH population models. IG, YB, and RK analyzed the simulation outputs and generated the key figures. RK developed the Python code used to compute kilonova light curves, based on literature prescriptions identified with input from MB. The code was further refined by IG.

\section*{Acknowledgements}
The authors acknowledge Wen-fai Fong, David Radice and Christopher Berry, for useful discussions, and Sebastiano Bernuzzi for clarifications regarding the employed fits.
IG acknowledges support from the Network for Neutrinos, Nuclear Astrophysics, and Symmetries (N3AS) Collaboration, NSF grant: PHY-2020275. IG also acknowledges the computational resources provided by the Gwave cluster, maintained by the Institute for Computational and Data Sciences at Penn State University, supported by NSF grants: OAC-2346596, OAC-2201445, OAC-2103662, OAC-2018299, and PHY-2110594. RK and YB acknowledge the support of Param Rudra, the high‑performance computing facility established under the National Supercomputing Mission at IIT Bombay. YB also acknowledges the
LIGO Lab computational resources, supported by NSF grants: PHY-0757058 and PHY-0823459. Parts of this work are included in YB's Master's thesis project at IIT Bombay. MB acknowledges support from the Eberly Research Fellowship at the Pennsylvania State University and the Simons Collaboration on Extreme Electrodynamics of Compact Sources (SCEECS) Postdoctoral Fellowship at the Wisconsin IceCube Particle Astrophysics Center (WIPAC), University of Wisconsin-Madison.

\section*{Data availability}
The data underlying this article is available in \href{https://doi.org/10.5281/zenodo.20752861}{Zenodo}~\citep{data}.

\bibliographystyle{aa}
\bibliography{references.bib}

\appendix
\nolinenumbers

\section{Detailed description of Kilonova modeling} \label{appsec:det_kn_mod}

\subsection{Fits from numerical relativity simulations}
\label{appsubsec:det_kn_mod_nr}

The matter content of the viscous or neutrino-driven wind and dynamical ejecta for BNS/NSBH mergers is determined by their binary parameters. Although numerical relativity simulations are required to accurately predict the ejecta masses for these mergers, here we adopt analytical expressions inferred from numerical simulations that cover a broad range of binary parameters.

\subsubsection{Fits for BNS mergers} \label{appsubsubsec:bns_fits}

For BNS mergers, the dynamical ejecta mass is estimated by \citep{Radice:2018pdn,Dietrich:2016fpt,Kawaguchi:2016ana}
\begin{equation}
\label{BNS_dyn}
\begin{aligned}
\frac{M_{\rm dyn}^{\rm BNS}}{10^{-3}M_{\odot}}
={}& \Biggl[
    \alpha \left(\frac{M_2}{M_1}\right)^{1/3}
    \left(\frac{1-2C_1}{C_1}\right)
    + \beta \left(\frac{M_2}{M_1}\right)^n
\\
&\qquad
    + \gamma \left(1-\frac{M_1}{M_{b,1}}\right)
    \Biggr] M_{b,1}
    + (1\leftrightarrow 2) + \delta .
\end{aligned}
\end{equation}
where $C_{i}=GM_i/(R_ic^2)$ is the NS compactness corresponding to gravitational mass $M_i$ and radius $R_i$, and $M_{b,i}$ is the baryonic mass of the NS. With the \texttt{M0RefSet + M0/M1Set} models with neutrino absorption and emission, \citet{Nedora:2020qtd} provides the fitting coefficients for $\log_{10} M_{\rm dyn}^{\rm BNS}$ as $\alpha=-0.1004$, $\beta=-0.4403$, $\gamma=-0.6452$, $\delta=0.2696$, and $n=0.3222$.

For the remnant disk mass, we use the analytical fit from \citet{Radice:2018pdn}, given by
\begin{equation}
\label{BNS_disk}
\log_{10}\left(\frac{M_{\rm disk}^{\rm BNS}}{M_{\odot}}\right) = {\rm max}\left[-3.0,\log_{10}\left(\alpha + \beta\tanh\left(\frac{\tilde{\Lambda}-\gamma}{\delta}\right)\right)\right],
\end{equation}
with $\alpha=0.1206$, $\beta=0.05095$, $\gamma=471.0$, and $\delta=0.5351$ \citep{Nedora:2020qtd}. Here, $\tilde{\Lambda}$ is the reduced tidal deformability parameter. $30\%$ of this mass is chosen to contribute to disk wind, via neutrino-driven winds and viscous ejecta components~\citep{Kashyap2019-eo}.

The total unbound ejecta is calculated as the sum of the dynamical ejecta and disk wind. Following the three-component model in ~\citet{Villar:2017wcc}, we attribute 20\% of the total ejecta to the red and blue components each, and the remaining 60\% to the purple component (cf. Section~\ref{sec:kn_modeling}).

The velocity of the ejecta is given by \citep{Radice:2018pdn,Dietrich:2016fpt},
\begin{equation}
\label{BNS_vel}
v/c = \left[\alpha\left(\frac{M_1}{M_2}\right)(1+\gamma C_1)\right] + (1\leftrightarrow 2) + \beta,
\end{equation}
with $\alpha=-0.5631$, $\beta=1.109$, and $\gamma=-1.186$ \citep{Nedora:2020qtd}.

We note that the conclusions of this work can depend sensitively on the fitting formulae used to estimate the ejecta masses. These fits can incur substantial errors and, in some cases, may be ill-conditioned \citep{Nedora:2020qtd}. For e.g., in the disk mass fit of Eq.~\ref{BNS_disk}, the best-fit value $\delta=0.5351$ implies that for nearly all values of $\tilde{\Lambda}$ differing from $\gamma=471$ by more than $\sim 2$, the argument of the hyperbolic tangent saturates, such that $\tanh[(\tilde{\Lambda}-\gamma)/\delta]\approx \pm 1$. As a result, the inferred disk masses for the BNS population cluster into two preferred branches. This feature propagates directly into the KN light curves and is responsible for the bimodality seen in the BNS post-peak decline distributions in Section~\ref{sec:comparison}.

To assess the sensitivity of our results to this modeling choice, we repeat the analysis using an alternative set of fitting formulae for the BNS dynamical and disk ejecta from \citet{Kruger:2020gig}. The dynamical ejecta mass is given by
\begin{equation} \label{new_BNS_dyn}
\frac{M^{\rm BNS}_{\mathrm{dyn}}}{10^{-3} M_\odot} = \left(
\frac{\alpha}{C_1}
+ \beta \left(\frac{M_{2}}{M_{1}}\right)^n
+ \gamma C_1
\right) M_1
+ (1 \leftrightarrow 2),
\end{equation}
where $\alpha=-1.261\times10^{-3}$, $\beta=1.449\times10^{-2}$, $\gamma=-4.715\times10^{-2}$, and $n=1.306$~\citep{Nedora:2020qtd}. The disk mass is estimated as
\begin{equation} \label{new_BNS_disk}
\log_{10}\left(\frac{M_{\rm disk}^{\rm BNS}}{M_{\odot}}\right) = \log_{10} M_1 + \max\left(-3.3,\,\gamma\log_{10}(\alpha C_{\rm NS,1}+\beta)\right),
\end{equation}
where $\alpha=-7.184$, $\beta=1.303$, and $\gamma=1.613$~\citep{Nedora:2020qtd}. Unlike Eq.~\ref{BNS_disk}, this prescription avoids the artificial saturation induced by the hyperbolic tangent. However, it predicts negligible disk ejecta for primaries with moderate to high compactness, approaching a floor of $M_{\rm disk}^{\rm BNS}\simeq 5\times10^{-4}\,\msun$ for $C_1\gtrsim 0.18$. This floor reflects the calibration uncertainty of the fit rather than a robust physical prediction for the disk mass~\citep{Kruger:2020gig}.

\begin{figure*}
    \centering
    \includegraphics[width=0.7\textwidth]{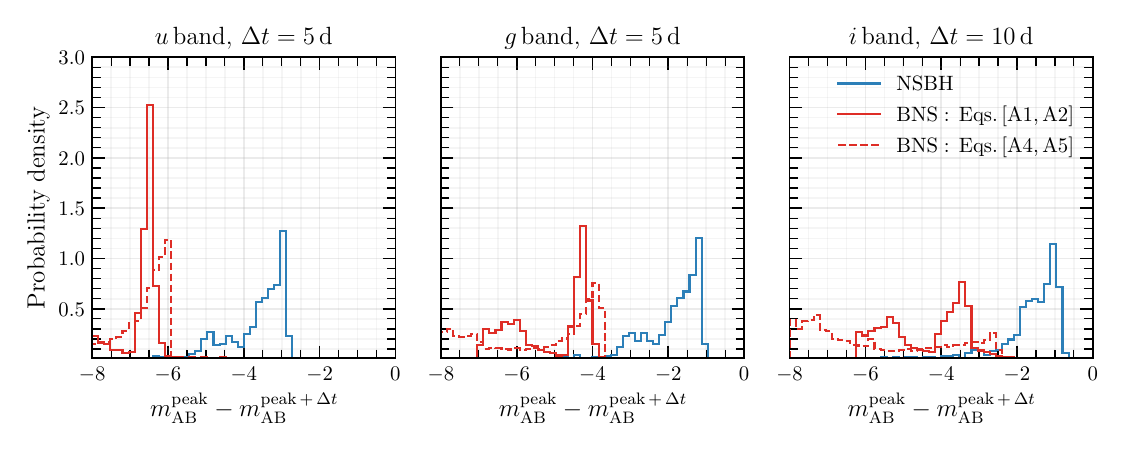}
  \caption{Comparison of the post-peak decline distributions obtained using two different prescriptions for the BNS ejecta masses, for the fiducial ``Uniform'' BNS and NSBH populations with the DD2 EOS. The alternative BNS fits in Eqs.~\ref{new_BNS_dyn} and \ref{new_BNS_disk} remove the bimodality introduced by the $\tanh$ disk mass fit used in the main text, but also impose an effective floor on the disk ejecta for sufficiently compact primary NSs. The new fits shift the BNS distribution toward slower-decaying light curves in all three bands. While the separation in the $u$ and $g$ bands remains fairly robust, the overlap between the BNS and NSBH distributions increases modestly in the $i$ band with the new fit.}
\label{appfig:new-bns-fit-comp}
\end{figure*}

Figure~\ref{appfig:new-bns-fit-comp} compares the fiducial ``Uniform'' BNS and NSBH populations, assuming the DD2 EOS, in the $u$, $g$, and $i$ band post-peak decline distributions when the alternative BNS ejecta fits of Eqs.~\ref{new_BNS_dyn} and \ref{new_BNS_disk} are used in place of the baseline prescription adopted in the main analysis. The new fits predict larger disk ejecta masses for low mass NSs, which shifts the BNS distributions rightward in the the three bands, toward more slowly declining KNe. This increases the overlap with the fiducial NSBH population, although a substantial fraction of NSBH mergers still occupies the more slowly evolving region.

The alternative fits avoid the tanh-saturation branch but introduce a different numerical-error-floor-driven branch for BNS systems with compact primary NS. The more rapidly declining branch comprises systems with $M_1 \gtrsim 1.6\,\msun$, for which the inferred disk mass is driven to the floor value of $\sim 5\times10^{-4}\,\msun$. Consequently, this branch is an artifact of the fit reaching its calibration floor rather than a robust physical prediction.

As expected, the quantitative degree of separation between BNS and NSBH KNe depends on the ejecta fits adopted. More generally, fits that predict larger BNS ejecta masses produce more slowly declining BNS light curves in the photometric bands. Because NSBH KNe are the more slowly declining population, this moves the two distributions closer together and reduces their separation in the three bands. On the other hand, fits that predict smaller BNS ejecta masses sharpen the contrast. Thus, the precise amount of overlap between the two merger classes is fit-dependent. Nevertheless, even under these alternative prescriptions, BNS and NSBH KNe remain reasonably separable in the regions of parameter space identified in this work.

\subsubsection{Fits for NSBH mergers} \label{appsubsubsec:nsbh_fits}

Depending on the initial BH and NS parameters, the outcome of an NSBH merger can be two-fold: either the NS plunges directly into the BH before it can be tidally disrupted, or the tidal forces on the NS become strong enough to disrupt it before it reaches ISCO. The binary parameters that determine this outcome are the binary mass ratio $q = M_{\rm BH}/M_{\rm NS}$ \citep{Kyutoku:2011vz,Foucart:2011mz}, BH spin magnitude $a_{\rm BH}$  and orientation \citep{Pannarale:2010vs,Foucart:2012nc}, and NS radius $R_{\rm NS}$, given by the EOS \citep{Duez:2009yy,Kyutoku:2010zd}.
NS tidal disruption by BH is facilitated by a larger $R_{\rm NS}$ (for smaller $C_{\rm NS}$), smaller $M_{\rm BH}$ and/or larger $\chi_{\rm BH}$. The amount of NS material ejected before falling into the BH is determined by the relative positions of the tidal disruption radius and the ISCO radius, where the latter is given by
\begin{equation}
\label{Risco}
\frac{R_{\rm ISCO}}{M_{\rm BH}} = 3 + Z_2 - {\rm sign}(\chi_{\rm BH})\sqrt{(3-Z_1)(3+Z_1+2Z_2)}
\end{equation}
where $Z_1 = 1 + (1 - \chi_{\rm BH}^2)^{1/3}[(1+\chi_{\rm BH})^{1/3} + (1-\chi_{\rm BH})^{1/3}]$ and $Z_2 = \sqrt{3\chi_{\rm BH}^2 + Z_1^2}$.

\citet{Kruger:2020gig} estimated the mass of NSBH dynamical ejecta with the analytical fit
\begin{equation} \label{eq:NSBH_dyn}
    \frac{M_{\rm dyn}^{\rm NSBH}}{M_{\rm b,NS}} = a_1\, q^{n_1} \left(\frac{1-2C_{\rm NS}}{C_{\rm NS}}\right) - a_2\, q^{n_2} \frac{R_{\rm ISCO}}{M_{\rm BH}}+a_3,
\end{equation}
where $a_1=0.007116$, $a_2=0.001436$, $a_3=-0.02762$, $n_1=0.8636$, and $n_2=1.6840$ are the best-fit parameters. Here, $M_{\rm b,NS}$ is the NS baryonic mass, and $C_{NS}$ is its compactness. Negative values obtained from equation (\ref{eq:NSBH_dyn}) represent no dynamical mass ejected post NSBH merger. \citet{Foucart:2016vxd} estimated the average velocity of dynamical ejecta as $v_{\rm dyn}^{\rm NSBH}=(0.0149 \,q+0.1493)c$.

\citet{Foucart:2018rjc} provides the following fit for the remaining baryonic mass outside the remnant BH $\sim 10$ ms after merger:
\begin{equation} \label{eq:Mejtot}
\frac{M_{\rm rem}^{\rm NSBH}}{M_{\rm b,NS}} = {\rm max}\left[\left(\alpha\frac{1-2C_{\rm NS}}{\eta^{1/3}} - \beta\frac{R_{\rm ISCO}}{M_{\rm BH}}\frac{C_{\rm NS}}{\eta} + \gamma\right)^{\delta},0\right]
\end{equation}
where $\alpha=0.406$, $\beta=0.139$, $\gamma=0.255$, $\delta=1.761$ are fit parameters, and $\eta=q/(1+q)^2$, also referred to as the symmetric mass ratio.
$M_{\rm dyn}^{\rm NSBH}$ and $M_{\rm rem}^{\rm NSBH}$ are found to be larger for a smaller $M_{\rm BH}$, larger $a_{\rm BH}$ and stiffer NS EOS (i.e., smaller $C_{\rm NS}$).
The mass of the NSBH accretion disk is estimated using $M_{\rm disk}^{\rm NSBH} = M_{\rm rem}^{\rm NSBH} - M_{\rm dyn}^{\rm NSBH}$. A significant portion of this disk can become gravitationally unbound through outflows that are either thermally or magnetically driven. The mass loss due to NSBH disk wind can be estimated as \citep{Fernandez:2020oow},
\begin{equation} \label{eq:Mejdyn}
\frac{M_{\rm wind}^{\rm NSBH}}{M_{\rm disk}^{\rm NSBH}} = \xi_1 + \frac{\xi_2 - \xi_1}{1 + e^{1.5(q-3)}},
\end{equation}
where we assume average values for the free parameters $\xi_1=0.18$ and $\xi_2=0.29$ \citep{Raaijmakers:2021slr}.

Unlike dynamical ejecta for NSBH mergers, winds originating from the remnant accretion disk are expected to be spherically symmetric. Although the velocity of the disk wind ejecta is quite uncertain, recent simulations have shown that the velocity for bulk of the material is centered around $\sim 0.1c$ \citep{Siegel:2017nub,De:2020jdt,Fernandez:2020oow}.

In Section~\ref{subsec:comp_modpars}, we also examine the effect of varying $Y_e$ in the dynamical and disk wind ejecta relative to the fiducial values $Y_e = 0.1$ and $0.3$, respectively. Following \citet{Ekanger:2023mde}, we consider $Y_e$ in the ranges $0.05-0.1$ for the dynamical ejecta and $0.1-0.5$ for the disk ejecta. Because the KN model takes the opacity as an input parameter, we map the assumed $Y_e$ values to $\kappa$ by fitting to the $Y_e-\kappa$ data of \citet{Tanaka:2019iqp}. Our best-fit relation is
\begin{equation} \label{eq:tanaka-fit}
\kappa(Y_e)=1.9469+\frac{24.4696}{0.6991+\left(4.6822\,Y_e\right)^{7.8073}}.
\end{equation}
This fit yields a smooth decrease in $\kappa$ to $\sim 2\ \mathrm{cm}^2\,\mathrm{g}^{-1}$ at $Y_e \sim 0.5$, while approaching $\kappa \sim 38\ \mathrm{cm}^2\,\mathrm{g}^{-1}$ for $Y_e \lesssim 0.05$. Under this mapping, $Y_e=\{0.05,\,0.1,\,0.3,\,0.5\}$ corresponds to $\kappa=\{36.95,\,36.82,\,3.59,\,1.98\}\ \mathrm{cm}^2\,\mathrm{g}^{-1}$, respectively.

\subsubsection{Obtaining bandwise light curves} \label{appsubsubsec:get_bandwise}
After obtaining the ejecta masses and opacities, the heating rate $\dot{Q}$, as a function of time, is given by~\citep{Korobkin:2012uy},
\begin{equation}
   \dot{Q} = 4\times10^{18} \times M_{ej} \times \left(\frac{1}{2} - \frac{1}{\pi}\tan^{-1}\left(\frac{t-t_0}{s}\right)\right)^{1.3}
\end{equation}
where $M_{ej}$ is the ejecta mass, $t_0 = 1.3$ and $s=0.11$. Together with the thermal efficiency, $\epsilon_{th}$, obtained from \citet{Barnes:2016umi}, we approximate the bolometric light curve as,
\begin{equation}
    L_{bol}(t) = 2 \exp\left(-\frac{t^2}{\tau^2}\right)\int_{t_i}^{t} \dot{Q}(t') \times \epsilon_{th}(t') \times \frac{t'}{\tau^2} \times \exp\left(\frac{{t'}^2}{\tau^2}\right)\,dt'
\end{equation}
where $\tau = \sqrt{\tau_{diff}\tau_{exp}}$ (c.f. Section~\ref{sec:kn_modeling}), and $t_i$ is the initial time. To estimate the bandwise light curves from the bolometric luminosity, we model the emission as a blackbody with photospheric temperature $T$ and radius $R$. The photosphere expands with the ejecta as $R = v_{ej}\times t$, where $v_{ej}$ is the velocity of the ejecta. $T$ and $R$ characterize the associated spectral flux density $f_{\nu}$:
\begin{equation}
    f_{\nu} = \frac{2h\nu^{3}}{c^2} \frac{1}{e^{\frac{h\nu}{kT}}-1}\, \left(\frac{R}{D_L}\right)^2,
\end{equation}
where $h$ is the Planck constant, $\nu$ is the frequency corresponding to the photometric band, $k$ is the Boltzmann constant, and $D_L$ is the luminosity distance of the system. Following \citet{Villar:2017wcc}, we impose a temperature floor $T_c=3000$~K to approximate the onset of recombination of elements to neutral~\citep{Barnes:2013wka}. Once the freely expanding ejecta cools below $T_c$, the photosphere radius recedes according to
\begin{equation}
    R\,(t)=\left(\frac{L_{\rm bol}(t)}{4\pi \sigma_{\rm SB} T_c^4}\right)^{1/2},
\end{equation}
where $\sigma_{\rm SB}$ is the Stefan-Boltzmann constant.
This captures the drop in opacity associated with the recombination. The color evolution remains constant in time, flattening the light curve, while the effective photosphere moves inward through the ejecta~\citep{Barnes:2013wka}. We note that, because the bluest bands lie on the Wien tail of the cooling photosphere, the predicted $u$ band flux is particularly sensitive to this temperature treatment. This motivates our use of the $g$ band as a more robust alternative in the main text.

The spectral flux can then be converted to AB magnitude $m_{\rm AB}$ using
\begin{equation}
    m_{\mathrm{AB}} = -2.5\,\mbox{log}_{10}\,f_{\nu}\,-\,48.6.
\end{equation}
For the $ugrizy$ bands, we use the following wavelengths: $\lambda_u = 3.546\times10^{-7}\,\mathrm{m}$, $\lambda_g = 4.670\times10^{-7}\,\mathrm{m}$, $\lambda_r = 6.156\times10^{-7}\,\mathrm{m}$, $\lambda_i = 7.472\times10^{-7}\,\mathrm{m}$, $\lambda_z = 8.917\times10^{-7}\,\mathrm{m}$, and $\lambda_y = 1.0305\times10^{-6}\,\mathrm{m}$.

\subsubsection{Example light curves and comparison with the DECam observation of GW170817's counterpart} \label{appsubsubsec:comp_gw170817}

\begin{figure}
    \centering
    \includegraphics[width=0.5\textwidth]{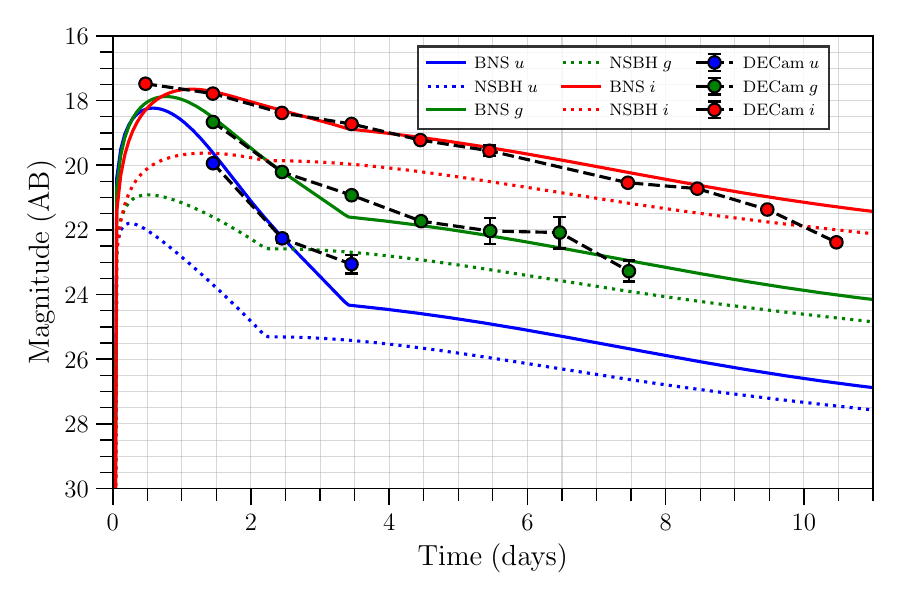}
    \caption{Example KN light curves in the $u$ (blue), $g$ (green), and $i$ (red) bands for a representative BNS (solid) and NSBH (dotted) system, computed with our semi-analytic framework and the DD2 EOS, compared with the DECam photometry of AT2017gfo, the optical counterpart of GW170817 (filled circles, as reposted by ~\citet{DES:2017kbs} and compiled by ~\citet{Cowperthwaite:2017dyu}). Both systems are placed at $40$ Mpc. The BNS adopts the GW170817-inferred masses
    $(m_1,m_2)=(1.4,1.35)\,\msun$, while the NSBH adopts $(m_{\rm BH},m_{\rm NS})=(3,1.35)\,\msun$ with $a_{\rm BH}=0$ and is shown only for contrast. The model reproduces the observed $g$ and $i$ band evolution well, whereas the $u$ band model declines faster than the sparse $u$ band data, consistent with the temperature-floor sensitivity discussed in Appendix~\ref{appsubsubsec:get_bandwise}.}
\label{appfig:comp-gw170817}
\end{figure}

Here we present example light curves computed with the framework described above and compare them with the observed KN counterpart of GW170817. For the BNS we
adopt the GW170817-inferred component masses $(m_1, m_2) = (1.4,1.35)\,\msun$
~\citep{LIGOScientific:2017vwq}, and for the NSBH we adopt
$(m_{\rm BH},m_{\rm NS})=(3,1.35)\,\msun$ with $a_{\rm BH} = 0$. We fix the EOS to DD2 and place both systems at a luminosity distance of $40$ Mpc. The NSBH system
is included only for contrast, while the DECam data are compared against the BNS model.

Figure~\ref{appfig:comp-gw170817} shows the resulting $u$, $g$, and $i$ band light curves together with the DECam observations of AT2017gfo~\citep{DES:2017kbs},
using the photometry compiled by~\citet{Cowperthwaite:2017dyu}. For this low mass,
non-spinning BH, the NSBH ejecta mass is small and its KN is correspondingly fainter than the BNS KN by $\gtrsim 2$ mag in all three bands, most strongly in the blue. In both systems, the bluer bands peak earlier and decline faster: the
$u$ band peaks first and then drops steeply, followed by $g$, while the redder $i$ band peaks later and fades most gradually.

\begin{figure*}[h!]
    \centering
    \begin{subfigure}{0.48\textwidth}
        \centering
        \includegraphics[width=\linewidth]{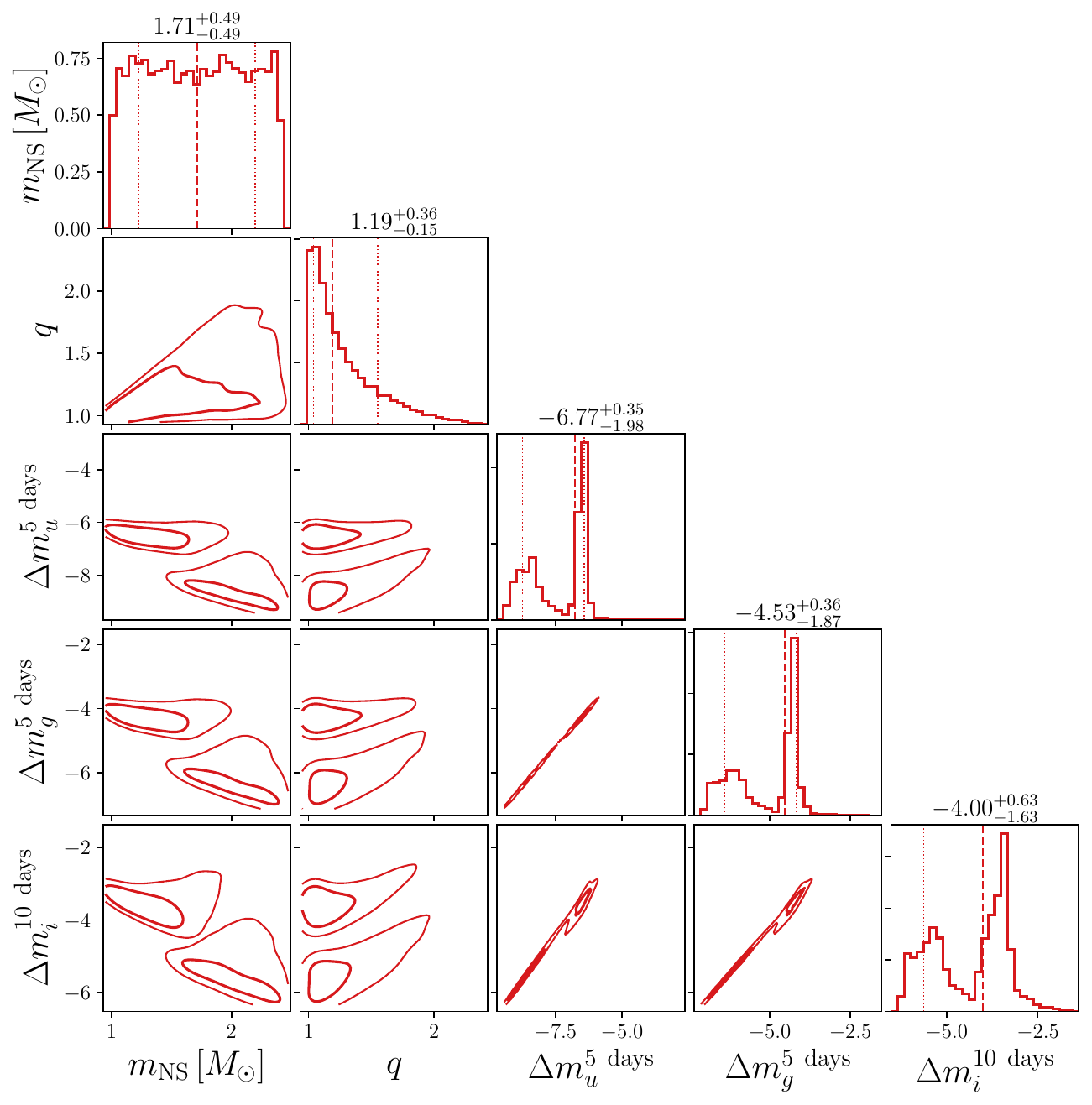}
    \end{subfigure}
    \hfill
    \begin{subfigure}{0.48\textwidth}
        \centering
        \includegraphics[width=\linewidth]{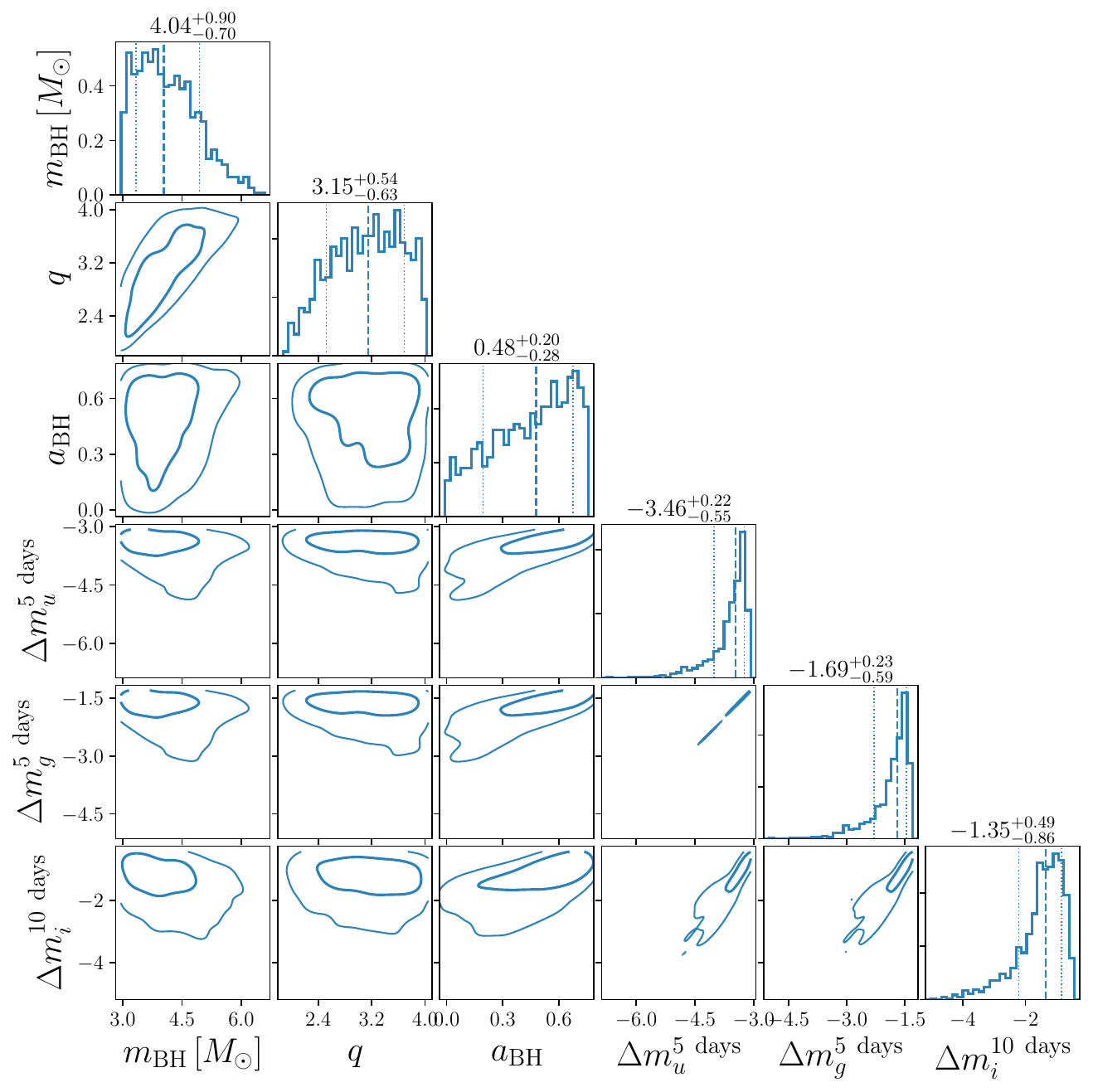}
    \end{subfigure}

    \caption{Dependence of the post-peak decline of KN light curves on the binary parameters for BNS (left) and NSBH (right) mergers. For BNS, the plot shows the relationship between the primary NS mass $m_{\rm NS}$, the binary mass ratio $q$, and the post-peak decline in the $u$, $g$ and $i$ bands. For NSBHs, we show the dependence of post-peak decline in the three bands with the mass of the BH $m_{\rm BH}$, $q$, and the dimensionless spin of the BH $a_{\rm BH}$. For both merger classes, results are shown for the fiducial ``Uniform'' population and the DD2 EOS. The plotted distributions illustrate how the decline rates in the $u$, $g$, and $i$ bands correlate with the binary masses and, for NSBH mergers, also with the BH spin.}
    \label{appfig:bin_corner}
\end{figure*}

The simulated BNS $g$ and $i$ band light curves reproduce the corresponding DECam measurements reasonably well, with appreciable deviations appearing only at late
times. The $u$ band, however, departs from the data already around $\sim 3$ d, where the model declines more steeply than the observed points. Although the $u$ band coverage of GW170817's counterpart is sparse, this suggests that our prescription may overpredict the late-time $u$ band decline, plausibly because the bluest band is the most sensitive to the blackbody and recombination temperature-floor treatment
(Appendix~\ref{appsubsubsec:get_bandwise}). This comparison provides direct, albeit
limited, observational support for treating the $u$ band decline with caution and for highlighting the $g$ band, which here tracks the data well, as a more robust
alternative diagnostic.

\section{Effect of binary parameters on ejecta properties} \label{appsec:bin_prop_ejec}

Figure~\ref{appfig:bin_corner} illustrates how the post-peak decline of the KN light curves depends on the underlying binary parameters. For BNS mergers, the left panel of the figure shows the relation between the primary NS mass $m_{\rm NS}$, the binary mass ratio $q$, and the post-peak decline in the $u$ band, $\Delta m_u^{\rm 5\,days} = m_{u, {\rm AB}}^{\rm peak}-m_{u, {\rm AB}}^{\rm peak+5\,{\rm d}}$, in the $g$ band, $\Delta m_g^{\rm 5\,days}$, and the $i$ band, $\Delta m_i^{\rm 10\,days}$. For NSBH mergers, we instead show how the same decline measures depend on the BH mass $m_{\rm BH}$, the mass ratio $q$, and the dimensionless BH spin $a_{\rm BH}$.

For BNS systems, lower mass NSs are less compact and more easily disrupted, yielding larger ejecta masses. This trend is reflected in the strong correlation between slowly declining $u$, $g$, and $i$ band light curves and low primary NS masses, $m_{\rm NS} \lesssim 1.5\,\msun$. Even when the primary NS is more massive, a sufficiently low mass companion can still lead to higher ejecta mass and produce slower post-peak evolution. This contributes to the trend with mass ratio, in which more asymmetric BNS mergers may also decline more slowly in the three bands. The same behavior is also consistent with numerical simulations showing that unequal-mass BNS mergers can produce more ejecta than nearly equal-mass systems~\citep{Camilletti:2022jms}.

For NSBH systems, the dominant parameters are the BH mass and spin. Smaller BH masses and larger aligned spins reduce $R_{\rm ISCO}$ and are conducive to tidal disruption, which in turn increases both the unbound ejecta and the remnant disk mass. This is clearly seen in the right panel of Figure~\ref{appfig:bin_corner}, where systems with lower $m_{\rm BH}$ (correspondingly, $q\lesssim 3$) and higher $a_{\rm BH}$, are associated with slower decline in both the $u$ and $i$ bands.

An important feature is that the post-peak decline remains well-separated even when the mass parameters of BNS and NSBH systems become comparable, for e.g., between a low mass BH with $m_{\rm BH}\sim 2.5\,\msun$ and a high mass NS with $m_{\rm NS}\sim 2.5\,\msun$. 
In our models, the high mass NS (BNS) configurations decline rapidly in the blue, with $\Delta m_u^{\rm 5\,days} \sim -9$ and $\Delta m_g^{\rm 5\,days} \sim -7$ while remaining comparatively fast in the red as well, $\Delta m_i^{\rm 10\,days} \sim -6$. The low-mass BH (NSBH) configurations give markedly different, more slowly declining values of $\Delta m_u^{\rm 5\,days} \sim -3$, $\Delta m_g^{\rm 5\,days} \sim -1.5$, and $\Delta m_i^{\rm 10\,days} \sim -1$. However, also note that numerical simulations in this region of parameter space are sparsely available, limiting the accuracy of the utilized remnant fits for such systems. At current and near-future GW detector sensitivities, such low mass BHs can be difficult to distinguish from high mass NSs using GW information alone~\citep{Golomb:2024mmt,Dhani2025-rq}. Thus, if the observed distinction persists the test of more complex simulations, the post-peak decline in photometric bands can provide an independent way to characterize such systems.

\section{Bandwise temporal evolution of KNe} \label{appsec:all_bands}

\begin{figure*}
    \centering
    \includegraphics[width=0.9\textwidth]{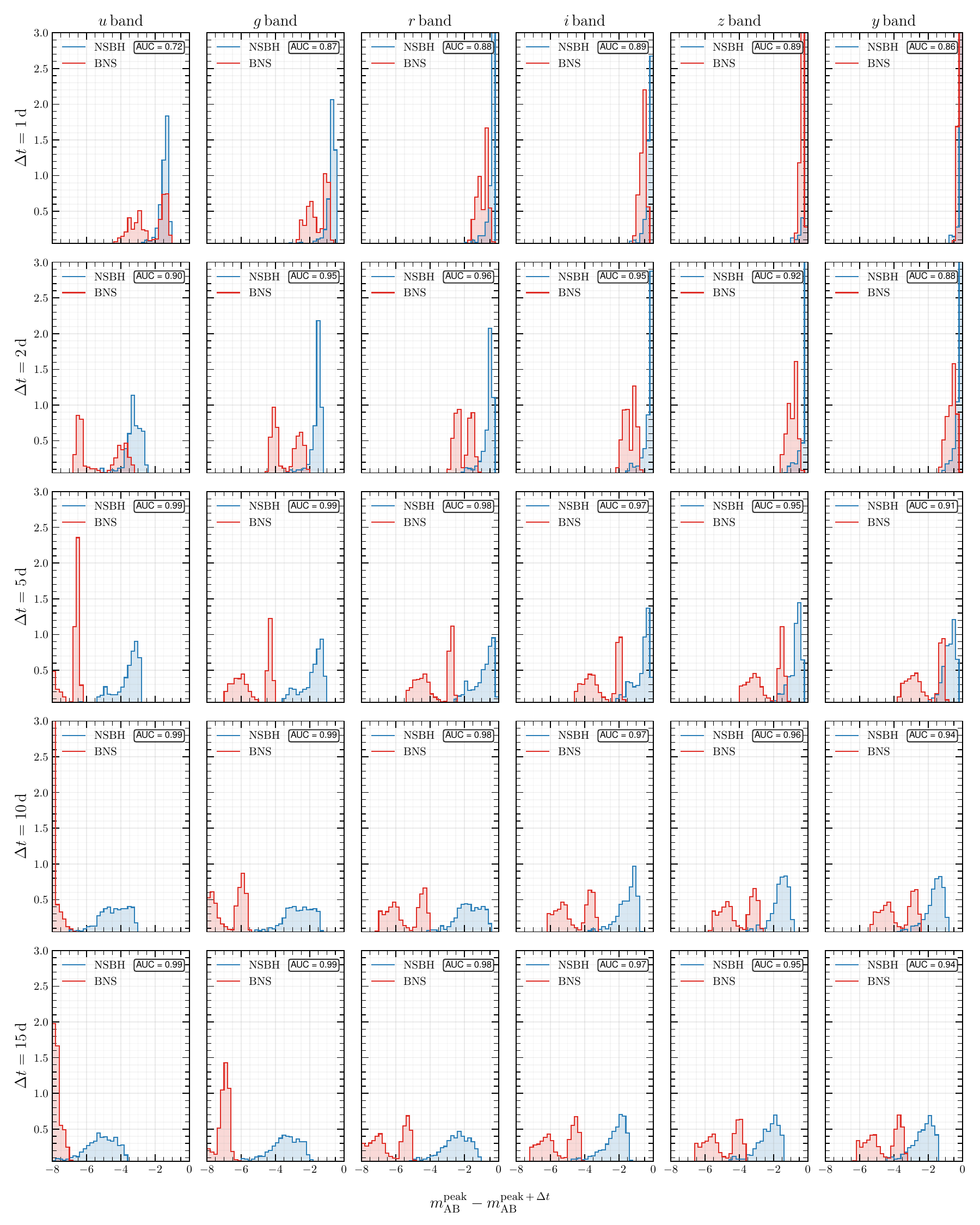}
  \caption{Post-peak decline distributions of BNS (red) and NSBH (blue) KN light curves in the optical $ugrizy$ bands, evaluated at $\Delta t = 1, 2, 5, 10$, and $15$\,d after peak, combining samples across all three NS EOS for the fiducial ``Uniform'' population. The AUC quoted in each panel measures the separation between the two classes (see Section~\ref{sec:comparison} and Figure~\ref{fig:auc_time}). The separation in the bluer bands ($u$, $g$) is weakest at early times and is largest at $\Delta t \simeq 5$\,d, while the reddest bands ($z$, $y$) reach their maximum separation around $\Delta t \simeq 10$\,d. These trends motivate the choice of the $u$ and $g$ bands at 5\,d after peak and the $i$ band at 10\,d after peak in the main text.}
\label{appfig:all_bands}
\end{figure*}

Figure~\ref{appfig:all_bands} shows the post-peak decline distributions in the $ugrizy$ bands for both BNS and NSBH KNe, evaluated 1, 2, 5, 10, and 15\,d after peak, with the corresponding AUC annotated in each panel.

To quantify the separation in each panel we use the AUC~\citep{HanleyMcNeil1982}, which has a simple operational meaning: it is the fraction of all BNS--NSBH pairs in which the NSBH event fades more slowly than its BNS counterpart. Concretely, for a given band and $\Delta t$, we compare every BNS decline $x^{\rm B}_i$ with every NSBH decline $x^{\rm N}_j$ (with $x \equiv m_{\rm AB}^{\rm peak}-m_{\rm AB}^{\rm peak+\Delta
t}$), count the pairs in which the NSBH fades more slowly ($x^{\rm N}_j > x^{\rm B}_i$), add half a count for every exact tie, and divide by the total number of pairs $n_{\rm B}\times n_{\rm N}$:
\begin{equation}
\label{eq:auc}
{\rm AUC} = \frac{1}{n_{\rm B}\, n_{\rm N}}\sum_{i=1}^{n_{\rm B}}\sum_{j=1}^{n_{\rm N}}
\left[\,\mathcal{I}\!\left(x^{\rm N}_j > x^{\rm B}_i\right)
+ \tfrac{1}{2}\,\mathcal{I}\!\left(x^{\rm N}_j = x^{\rm B}_i\right)\right],
\end{equation}
where $\mathcal{I}(\cdot)$ equals $1$ when the condition inside it is true and $0$
otherwise, so the double sum simply tallies the favorable pairs. This quantity is the normalized Mann--Whitney $U$ statistic~\citep{MannWhitney1947}, and equals the area under the
receiver-operating-characteristic curve, i.e., the AUC. Equivalently, it is the probability that a randomly chosen NSBH KN fades more slowly than a randomly chosen BNS KN. Since it counts only how often one class outranks the other, the AUC is set entirely by the relative ordering of the two samples. It is insensitive to the detailed shape
of either distribution and to the assumed relative rates of BNS and NSBH mergers. As stated in Section~\ref{sec:comparison}, ${\rm AUC}=0.5$ corresponds to indistinguishable populations and ${\rm AUC}=1$ to perfect separation. We include only systems that produce a measurable KN in the band considered, excluding NSBH
binaries that do not disrupt and any events with undefined decline, so the quoted values are conditional on a detectable counterpart existing.

The bluer bands ($u$, $g$) show the clearest separation between the two merger classes at $\Delta t \simeq 5$\,d, where their AUC reaches $\gtrsim 0.99$. At earlier times $(\Delta t \lesssim 2\,{\rm d})$ they are in fact the weakest discriminators, since the rapidly fading BNS blue component has not yet pulled the two populations apart. As one moves to redder bands, the maximum separation occurs progressively later, with the reddest bands ($z$, $y$) peaking around $\Delta t \simeq 10$\,d. Motivated by these trends, in the main text we use the $u$ band decline 5 days after peak and the $i$ band decline 10 days after peak as representative observables, with the $g$ band offered as a more robust alternative to $u$. More broadly, this behavior suggests that the separation between the two binary classes is a robust feature of the post-peak decline across the optical bands, rather than of any single band or epoch.

\end{document}